# Statistical Significance and Effect Sizes of Differences among Research Universities at the Level of Nations and Worldwide based on the Leiden Rankings

Loet Leydesdorff,*[a] Lutz Bornmann,[b] and John Mingers[c]

**Abstract**

The Leiden Rankings can be used for grouping research universities by considering universities which are not statistically significantly different as homogeneous sets. The groups and intergroup relations can be analyzed and visualized using tools from network analysis. Using the so-called "excellence indicator" $PP_{\text{top-10\%}}$—the proportion of the top-10% most-highly-cited papers assigned to a university—we pursue a classification using (*i*) overlapping stability intervals, (*ii*) statistical-significance tests, and (*iii*) effect sizes of differences among 902 universities in 54 countries; we focus on the UK, Germany, Brazil, and the USA as national examples. Although the groupings remain largely the same using different statistical significance levels or overlapping stability intervals, these classifications are uncorrelated with those based on effect sizes. Effect sizes for the differences between universities are small ($w$ <.2). The more detailed analysis of universities at the country level suggests that distinctions beyond three or perhaps four groups of universities (high, middle, low) may not be meaningful. Given similar institutional incentives, isomorphism within each eco-system of universities should not be underestimated. Our results suggest that networks based on overlapping stability intervals can provide a first impression of the relevant groupings among universities. However, the clusters are not well-defined divisions between groups of universities.

[a] * corresponding author; Amsterdam School of Communication Research (ASCoR), University of Amsterdam PO Box 15793, 1001 NG Amsterdam, The Netherlands; loet@leydesdorff.net
[b] Division for Science and Innovation Studies, Administrative Headquarters of the Max Planck Society, Hofgartenstr. 8, 80539 Munich, Germany; bornmann@gv.mpg.de
[c] Kent Business School, University of Kent, Canterbury CT7 2PE, United Kingdom; j.mingers@kent.ac.uk

## 1. Introduction

Following the introduction of the "Shanghai rankings" of universities in 2004 (Academic Ranking of World Universities, ARWU, 2004), a quasi-industry of university rankings has emerged (e.g., Shin, Toutkoushian, & Teichler, 2011). The various rankings (for example, the Times Higher Education World University Rankings and QS World Universities Rankings) use somewhat different parameters such as quality of education, number of Nobel Prizes, number of articles in top-journals—however defined—or also the visibility of a university on the internet (e.g., Aguillo, Ortega, & Fernández, 2008; Harzing & Mijnhardt, 2015; Tang & Thelwall, 2004). Although there is some consensus about a group of most-elite universities, differing parameters and models may have considerable effects on lower-ranked universities. From this perspective, the reliability of rankings is low. Gingras (2016, at p. 75), for example, argued "that annual rankings of universities, be they based on surveys, bibliometrics, or webometrics, have no foundation in methodology and can only be explained as marketing strategies on the part of the producers of these rankings."

Are there significant differences among leading research universities, or are there homogeneous classes with no significant differences among them? European governments, for example, have funded state universities hitherto often using a scheme which assumes equality among them. But is this empirically the case? Martin (2010) argued that under market pressure inequalities can be expected to have increased in recent decades. Using a sample of 500 universities, however, Halffman & Leydesdorff (2010) showed that the Gini coefficient (a measure of inequality) of the distribution of publications over universities declined during the period 2003-2007. Using a similar methodology, Ville *et al.* (2006) found decreasing inequality in research outputs among Australian universities during the period 1992-2003. The authors suggest that "institutional isomorphism" has led to imitative behavior (DiMaggio & Powell, 1983). In other words, universities appear to have become more similar.

Universities are positioned similarly on the market and share the same incentives; lower-ranked universities tend to imitate innovations by the leading universities. One can also raise the methodological question of whether differences in the ratings are statistically significant or mainly an artifact of considering too many decimals, as in the case of differences among journals measured by the journal impact factor (which includes three decimals). Using detailed rankings, an impression of differences can be generated whereas equality may prevail (Waltman, 2016).

Among the university rankings, the so-called "Leiden Ranking of research universities" (LR; available at http://www.leidenranking.com/) stands out for its clarity about limitations, methodological care, and transparency (e.g., Frenken, Heimeriks, & Hoekman, 2017). LR is based on data of the Web-of-Science (WoS) of Clarivate Analytics which are processed by the Centre for Science & Technology Studies (CWTS) at Leiden University. After an initial phase, the methodology was firmly established at the time of the Leiden Ranking 2013 (Waltman *et al.*, 2012). The user of LR can interactively (on the internet) select a world region, country or discipline, and a preferred parameter to be used for the ranking. Furthermore, one can download the complete data for each year as Excel files.

In this study, we use the dataset of LR 2017 to explore the question of classifying universities using statistics. In addition to the numbers of publications, LR provides the numbers of publications in the top-10% segment of most highly-cited publications both as a number ($P_{top\ 10\%}$) and as a percentage ($PP_{top\ 10\%}$) normalized against the respective reference sets. Normalization against reference sets is needed because citation intensity varies among disciplines (Garfield, 1979; Moed, 2010). The reference sets, however, are themselves dynamic, making it difficult to compare results over a series of years. For this reason, all previous years are recalculated in LR using the model of the current year (Leydesdorff, Wouters, & Bornmann, 2016, pp. 2144f.).

$PP_{top\ 10\%}$ is a proportion and the differences among universities can therefore statistically be tested for significance using, for example, the *z*-test. Echoing Schneider's (2013; 2015) criticism of the use of significance tests in bibliometrics, however, Waltman (2016)—one of the conceptual organizers of the LR—argued against the use of both significance testing and confidence intervals based on boot-strapping, stating that "it seems likely that the use of statistical inference will lead to confusion and misunderstandings." In a section about "Responsible Use" on the website of the Leiden Rankings (at http://www.leidenranking.com/information/responsibleuse), the authors state:

> To some extent it may be possible to quantify uncertainty in university rankings (e.g., using stability intervals in the Leiden Ranking), but to a large extent one needs to make an intuitive assessment of this uncertainty. In practice, this means that it is best not to pay attention to small performance differences between universities. Likewise, minor fluctuations in the performance of a university over time can best be ignored. The focus instead should be on structural patterns emerging from time trends.

In our opinion, this advice begs the question, since one has no tools other than statistics to distinguish between differences among groups and variation within groups. Without statistics, the criteria would be subjective or the interpretation based on intuitive "rules of thumb" (Van Raan, 2005, at p. 7). However, LR offers "stability intervals" of the $PP_{top\ 10\%}$ (Lunneborg, 2000; Colliander & Ahlgren 2011, at p. 105). As Colliander & Ahlgren (2011, at p. 105) formulate: "[I]f two departments have overlapping stability intervals this indicates that there is no substantial difference between these departments." A stable result is one that is not influenced by including or excluding specific cases in the analysis. Stability intervals thus provide us with a second means to group universities.[1]

The critique of the use of inferential statistics finds its origins in the work of Cohen (1977, 1994) who proposed the use of "power analysis" as an alternative to significance testing. Statistical significance can be an effect of sample size and does not inform us about the strength of a relationship. Furthermore, significance testing assumes the specification of a null-hypothesis which can be tested against a sample. Our data does not allow for this.

The persistent (mis)use of statistical-significance testing in the literature prompted the American Statstical Association to issue a consensus statement in 2016 in which the use of null-hypothesis statistical testing (NHST) was strongly disapproved as a measure of evidence: "The *p*-value is a statement about data in relation to a specified hypothetical explanation, and is not a statement about the explanation itself. […] Any effect, no matter how tiny, can produce a small *p*-value if the sample size or measurement precision is high enough, and large effects may produce unimpressive *p*-values if the sample size is small or measurements are imprecise" (Wasserstein & Lazar, 2016, at pp. 131f.). Statistical-significance tests should therefore be accompanied by effect sizes as measures of "practical significance."

---

[1] Bornmann *et al.*'s (2013) analysis of LR 2011 compared the stability intervals with other possible ways to calculate standard errors (e.g., the standard errors of a binary probability). They found a perfect correlation between stability intervals and these other possible ways which are based on less data-intensive computing procedures.

Cumming (2013), for example, argued that empirical studies using inferential statistics should report not only statistical significance but also effect sizes. In his opinion, meaningful differences between groups of entities can only be uncovered if *both* test statistics are considered. In this study, we juxtapose the results of using stability intervals, significance-testing, and effect sizes in order to address the questions raised, such as whether groups of universities are not substantially different in terms of their rankings despite the impression of differences in their positions provided by these rankings. Whereas significance testing and stability intervals focus on inferences beyond the sample, effect sizes reflect the magnitude of differences.

Among the various measures of effect sizes, Cohen's **w** fits our type of data. Unlike the effect-size measures of differences between means and proportions (Cohen's ***d***; Cumming & Calin-Jageman, 2016), **w** is non-parametric and based on chi-square statistics. Note that $PP_{top10\%}$ is non-parametric. The application of ***w***, therefore, is straightforward, but the interpretation of effect sizes remains specific to the hypothetical model and the research design. While a statistical-significance level of 5% is defined as a cut-off at $z = 1.96$, **w** has an indicative interpretation: **w** = .10 can be considered as "small;" **w** = .30 as "medium;" and **w** = .50 as a "large" effect (Cohen, 1988, at p. 227). However, Cohen (1988, at p. 226) "reiterates a word of caution about the use of constant **w** values to define a given level of departure, such as the operational definitions of "small," "medium," and "large" […]." He warns (at p. 20) that "[t]he absolute size of a point is a consequence of arbitrariness in the decision by the investigator, and/or in the scale construction method, and/or the writing or selection of the items."

In summary, one cannot expect the measurement and analysis to provide us with clear-cut answers to the question of how many groups are to be distinguished, but the combined assessment in terms of these tests can inform us about the fit or lack of fit between the results and the model assumption that universities can be ranked meaningfully because they are sufficiently different. If groups are distinguishable, however, the borderlines among them remain disputable. Our objective is to discuss how one might address the research question of how and whether to group universities into classes. This question is highly policy-relevant, since governments for example may wish to differentiate policies for different classes of universities. We use tools from network analysis to illustrate our results.

With a similar objective, Bornmann & Glänzel (2017) used the method of Characteristic Scores and Scales (CSS; Glänzel, 2007) to group universities into meaningful classes (poorly cited, fairly cited, remarkably, and outstandingly cited). However, these authors did not apply statistics beyond the description of publication data and did not visualize the results which facilitates understanding of the results. Using network analysis, we decompose the university groups. Thus, the fuzziness or clarity of the distinctions can also be estimated using, for example, modularity *Q* (Newman & Girvan, 2004; Blondel *et al*., 1988).

## 2. Statistics

### *2.1. Stability intervals*

The construction of stability intervals in LR is based on bootstrapping (Waltman *et al*., 2012, at p. 2429): one thousand samples are drawn from each university's set of publications, leading to thousand $PP_{top\ 10\%}$. In order to obtain a 95% stability interval, the authors take the 2.5$^{th}$ and the 97.5$^{th}$ percentiles of the $PP_{top\ 10\%}$ distribution (based on the samples) as the lower and upper bounds of the stability intervals.

When the stability intervals of two universities overlap, the distinction between them in terms of the indicator can be rejected (Colliander & Ahlgren, 2011). Since each of the two universities may be indistinguishable from other universities, we thus obtain so-called "weak" components in terms of network analysis. If both the upper and lower bounds of university A are contained within the stability interval of university B, the performance of the former can be seen as similar to the latter. In this case, we have a strong component since both arcs are present. We evaluate relations in terms of the arcs.

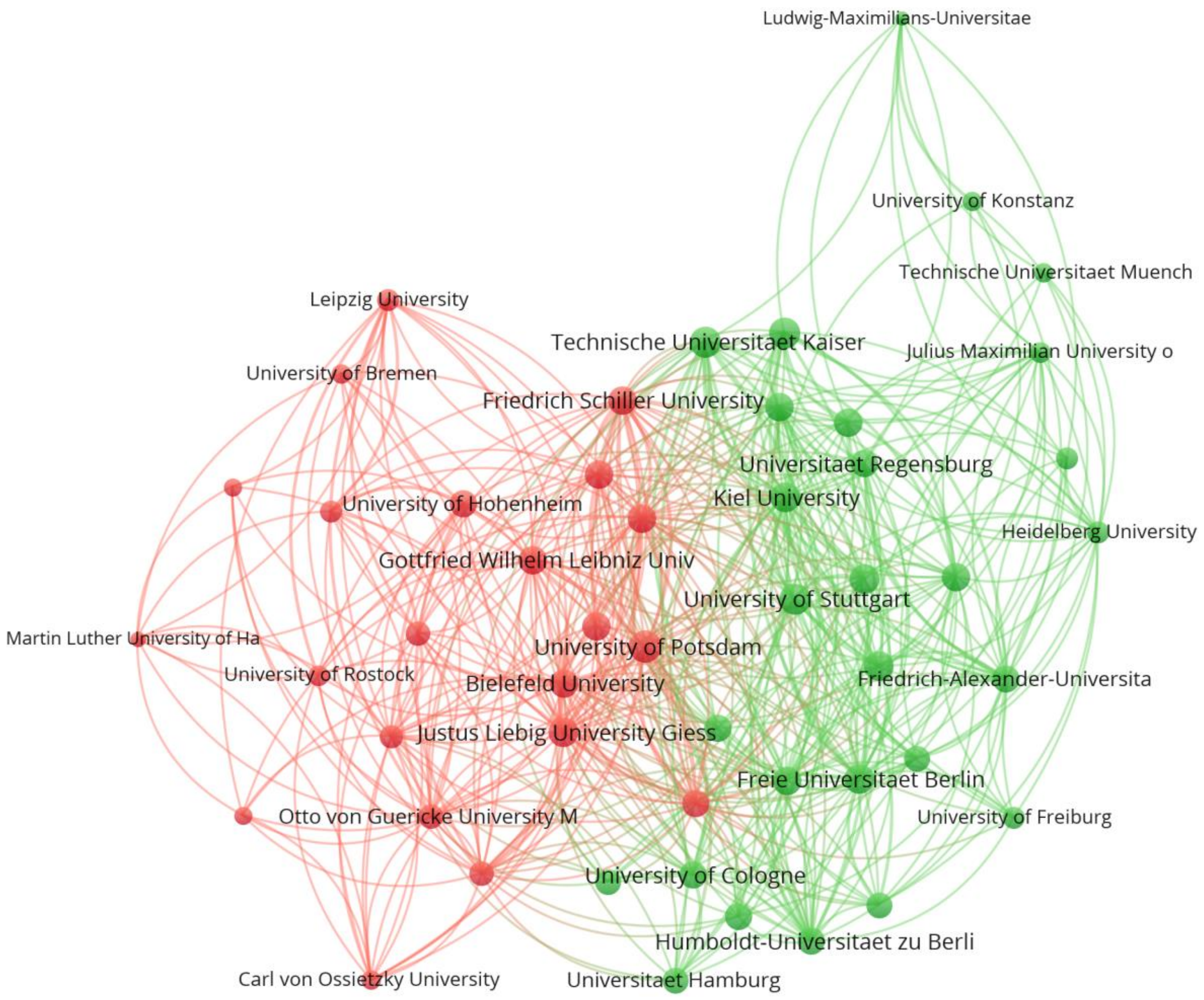

**Figure 1**: Overlapping stability intervals used as a grouping criterion for 50 German universities covered by the Leiden Ranking; two communities distinguished with modularity $Q = 0.21$ (Blondel *et al.*, 2008); VOSviewer used for the visualization.

As an example, Figure 1 provides a visualization of the two components which can thus be distinguished among fifty German universities covered by LR. The green group on the right side consists of established universities, while the red-colored group on the left side is populated with more marginal and recently founded universities. The vertical axis suggests a north-south tendency between "Munich" at the top and "Berlin" at the bottom.

Throughout this study, we use the Louvain-algorithm for community finding (Blondel *et al.*, 2008), because it provides less isolates than the algorithm of VOSViewer. In the case of Figure 1, for example, the community-finding algorithm of VOSViewer distinguishes a subgroup of "Berlin" universities within the group at the right, and the Heinrich-Heine Universität in Duesseldorf would be considered as a separate grouping at the interface between the two larger groups. Note that we use VOSviewer for the visualization (cf. Abramo, d'Angelo, & Grilli, 2016), but not for the decomposition.

*2.2. The z-test for estimating statistical significance*

The analysis based on stability intervals provides us with a binary value and does not exploit the further information of the indicator values. However, significance testing and effect sizes enable us to make these next steps. The *z*-test can be used to measure the extent to which an observed proportion differs significantly from expectation—in the case of $PP_{top\ 10\%}$, this is 10%—and whether the proportions for two institutions are significantly different. The test statistics can be formulated as follows:

$$z = \frac{p_1 - p_2}{\sqrt{p(1-p)\left[\frac{1}{n_1} + \frac{1}{n_2}\right]}} \qquad (1)$$

where: $n_1$ and $n_2$ are the numbers of all the papers published by institutions 1 and 2 (under the column "*P*" in the LR); and $p_1$ and $p_2$ are the values of $PP_{top\ 10\%}$ of institutions 1 and 2. The pooled estimate for proportion (*p*) is defined as:

$$p = \frac{t_1 + t_2}{n_1 + n_2} \qquad (2)$$

where: $t_1$ and $t_2$ are the numbers of top-10% papers of institutions 1 and 2. These numbers can be calculated on the basis of "*P*" and "$PP_{top\ 10\%}$". When testing values for a single university, $n_1 = n_2$, $p_1$ is the value of the $PP_{top\ 10\%}$, $p_2 = 0.1$, and $t_2 = 0.1 * n_2$ (that is, the expected number in the top-10%).

An absolute value of *z* larger than 1.96 indicates statistical significance of the difference between two ratings at the five percent level ($p < 0.05$). The threshold value for a test at the one percent level ($p < 0.01$) is 2.576; $|z| > 3.29$ for $p < 0.001$. In a series of tests for many institutions, one may wish to avoid family-wise accumulation of Type-I errors by using the Bonferroni correction; that is, $p_{\text{Bonferroni}} = \alpha / n$ where α is the original test-statistics and *n* the number of comparisons.

Universities which are not statistically significantly different can again be considered as belonging to the same performance group. Despite differences in $PP_{top\ 10\%}$ the performance of these universities can be denoted as similar in statistical terms. As above, this group membership is represented as links, so that groups can be visualized and analyzed using network software.

*2.3. Effect sizes*

In analogy to the *z*-test, 2 * 2 contingency tables are generated by comparing two universities in terms of their numbers of papers in the top-10% most highly-cited reference groups versus the other papers of each of the universities. The expectation is that universities are not different from each other or from the 10% level of $PP_{top\ 10\%}$. The larger the effect size, the more a difference is indicated. A measure of effect size ***w*** can be derived from chi-square statistics and is formalized as follows (Cohen, 1988, p. 216):

$$\boldsymbol{w} = \sqrt{\sum_{i=1}^{m} \frac{(p_{1i} - p_{0i})^2}{p_{0i}}} \tag{3}$$

where $p_{0i}$ is the proportion in cell *i* posited by the null hypothesis, and $p_{1i}$ the proportion in cell *i* posited by the alternative hypothesis; *m* reflects the number of cells. Note that the formula is similar to that for $\sqrt{\chi^2}$ except that relative values (proportions) are used instead of absolute values.

*2.4. Numerical example*

At http://www.leydesdorff.net/leiden11/index.htm the user can retrieve a file leiden11.xls which allows for feeding *P* and $PP_{top\ 10\%}$ values harvested from the LR for each two universities. The spreadsheet provides the significance level of the difference measured as both a *z*-score and a **w**-value. For example, Leiden University is listed (in the category "All sciences" of LR 2017) with *P* = 6,368 articles of which 13.8% participate in the top-10% layer for the comparable set worldwide ($PP_{top\ 10\%}$); the upper and lower bounds are 13.10 and 14.50. The University of Amsterdam has 8,519 articles with $PP_{top\ 10\%}$ = 14.5%, bounded between 13.90 and 15.10. The stability intervals are thus intersecting.

On the basis of this data, one can write the following contingency tables and derive the values of $\chi^2$ and ***w***:

| *observed values* | *top-10%* | *non-top* | | *proportions* | | |
|---|---|---|---|---|---|---|
| *Leiden* | 878.784 | 5489.216 | 6368 | 0.059030295 | 0.368725 | 0.42775576 |
| *Amsterdam* | 1235.255 | 7283.745 | 8519 | 0.082975415 | 0.489269 | 0.57224424 |
| | 2114.039 | 12772.96 | 14887 | 0.14200571 | 0.857994 | 1 |
| *expected values* | | | | | | |
| *Leiden* | 904.2924 | 5463.708 | 6368 | 0.06074376 | 0.367012 | 0.42775576 |
| *Amsterdam* | 1209.747 | 7309.253 | 8519 | 0.081261949 | 0.490982 | 0.57224424 |
| *contributions to the chi-square* | | | | *contributions to the effect size* | | |
| *Leiden* | 0.719542 | 0.119091 | 0.838633 | 0.000048 | 0.000008 | 0.000056 |
| *Amsterdam* | 0.537862 | 0.089021 | 0.626883 | 0.000036 | 0.000006 | 0.000042 |
| | | $\chi^2$= | 1.465 | | ***w*** = | Sqrt(0.000098) =0.099 |

**Table 1**: Numerical example of the computation of $\chi^2$ and Cohen's ***w***.

For the *z*-test one needs the pooled estimate. Using the values in the top panel of Table 1, *p* = (878.784 + 1235.255) / (6368 + 8519) = 0.1420. Using Equation 1 (above), it follows that |*z*| = 1.211. The difference between Leiden and Amsterdam is therefore not statistically significant and the effect size is small (***w*** < 0.1). In sum, the two universities can be considered as belonging to the same group.

## 3. Data

Data of the full LR sets can be downloaded in Excel format at http://www.leidenranking.com/downloads. LR 2017 analyzes 902 universities from 54 countries. The file contains ranks for these universities in the preceding years (in intervals of four years). Rankings are counted both fractionally and in whole numbers. Data is provided for "All sciences" and five major fields: (*i*) biomedical and health sciences, (*ii*) life and earth sciences, (*iii*) mathematics and computer science, (*iv*) physical sciences and engineering, (*v*) social sciences and humanities. We limit the analysis here pragmatically to "All sciences" (cf. Strotmann & Zhao, 2015), the last available period (2012-2015), and fractional counting. We include only the 10,898 fully-covered core journals and not the 3,900 non-core journals. However, the analysis can be repeated analogously using any subset and with other parameters.

For our purposes, we reorganize the file so that the fields "university," "country," "field," "period" (publication years), "fractional"(fractional or full counting of publications), "p" (number of publications), "p_top10," "pp_top10" and its upper and lower bounds are saved as a comma-separated data file. A dedicated routine (available at http://www.leydesdorff.net/software/leiden; see Appendix 1) reads this file as input and generates, for each country and the whole set, the following files:

1. A file in the Pajek format with universities as vertices and $z$-values as links insofar as $z < 2.576$ (the cutoff for $p < .01$). This file is named with the country name (e.g., "Germany.net" describing 50 German universities represented in the data). Files are thus generated for the 54 individual countries, and one additional file "world.net" contains the data for all 902 universities.

2. A second file in the Pajek format with similar information, but with **w-** values for the links; in this case, no threshold is set *a priori*. Each of these files has the same name as under 1, but "_w" is added to the country name as a root (e.g., "Germany_w.net").
3. A file in the Pajek format with similar information, but with value 1 for the links between universities with overlapping stability intervals, and 2 for the strong components. Each of these files has the same name as under 1, but "_o" is added to the country name as a root.
4. The *z*-values for testing the universities at the nodes against the 10% value of most highly-cited publications are stored in a file with the same country names, but with the extension ".vec". Since most network programs can handle only positive values, negative values of *z* are set equal to zero.
5. The full set of *z* values is retrievable from the Pajek files in the header, indicating the size of each node. Using these files, positive values can be represented in the visualization (using Pajek) by a circle and negative ones by a diamond. A partition file with the extension ".clu" for each country is generated distinguishing between positive and negative values of *z* (using "2" and "1", respectively.)

The Pajek (.net) format provides a kind of currency among programs for network analysis and visualization. We store the resulting measures (overlap, *z,* or ***w***) in the edge value between each two universities. Note that these universities are not necessarily related for example by citation or co-authorship. The use of network statistics is in this sense metaphorical. However, both VOSviewer and MDS (e.g., in NetDraw or SPSS) are based on showing structural similarities (in a vector space), in our case similar or different institutional impact performances in terms of statistical and practical significance. In VOSviewer, network links can additionally facilitate the interpretation.

## 4. Results

### *4.1. All universities in the Leiden Ranking 2017 (n = 902)*

Nine hundred of the 902 universities in the LR are linked into the largest component on the basis of overlaps in the stability intervals. The two exceptions are MIT and Rockefeller University. Although this may not come as a surprise in the case of MIT, 31.2% of the publications of Rockefeller University are part of the group of top-10% most-highly-cited papers. For MIT, this percentage is 25.0% and for Harvard 22.5%. However, the stability interval of Harvard intersects with that of Stanford, the University of California at Berkeley, the London School of Hygiene and Tropical Medicine, and Princeton University. Otherwise, four components are distinguished (Figure 2).

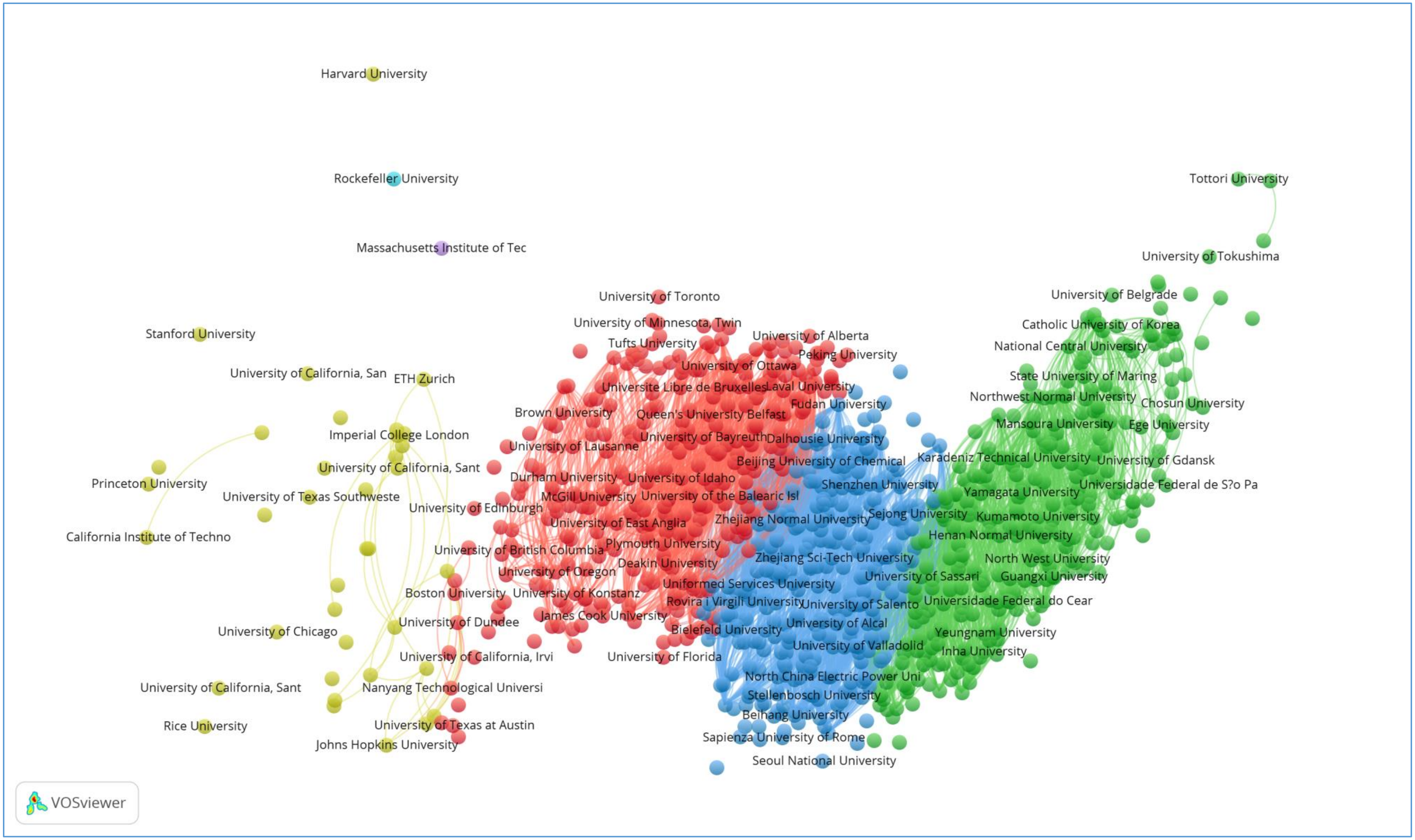

**Figure 2**: Clustering of 902 universities covered by the Leiden Ranking 2017 based on overlapping stability intervals; modularity $Q = 0.21$ (Blondel *et al*., 2008); VOSviewer used for the visualization; layout according to Kamada & Kawai (1989).

In the left-most component (yellow), one finds the leading American research universities and a few European and Asian ones. The second component (red) contains other American universities, European, Canadian, and some Chinese universities. The third component (blue) is dominated by Asian universities, but some large European universities such as La Sapienza in Rome are also positioned here. These universities are sometimes education-oriented with less emphasis on research. The final component (green) is composed of universities with a low track record in terms of research. The $PP_{top\ 10\%}$ of these universities is often below the expectation of 10%.

Given the large number of observations ($n$ = 902, therefore the number of possible comparisons is [902 * 901 /2 =] 406,351) a level of 1% is indicated for testing statistical significance. Figure 3 shows the separation of the 902 universities in LR into three groups plus the same two isolates. The grouping is based on the Louvain algorithm (Blondel *et al.*, 2008) and the structuring on spring-embedding the resulting network using Kamada & Kawai (1989). Node sizes are proportionate to the $z$ values for individual universities compared to the baseline of $PP_{top\ 10\%}$ = 10%.

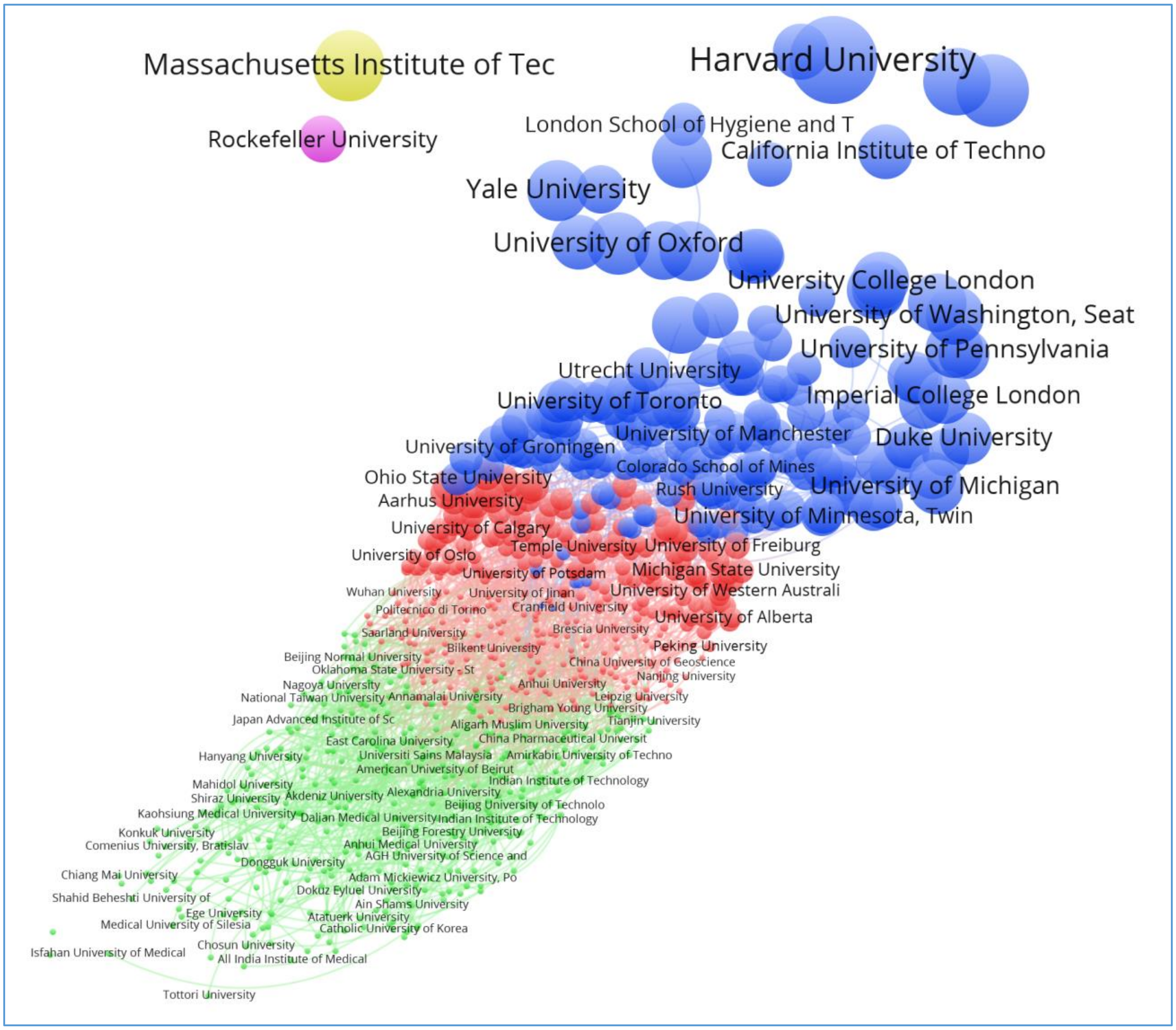

**Figure 3**: Three groups of universities are distinguished using $|z| > 2.576$; $p < .01$. Furthermore, two universities are isolates (MIT and Rockefeller University); $Q = 0.27$ (Blondel *et al*., (2008); layout according to Kamada & Kawai (1989).

Lowering the significance level (figure not shown) leads in this design to more groups because values of links above the respective $|z|$-values have been deleted. The main difference at the 5% level is the appearance of a fine structure at the top: a small group of leading American universities is distinguished from leading European universities (Oxford, Cambridge, ETH

Zurich) which were integrated with this top-group when values of *z* between 1.96 and 2.576 were first included.

In Figure 4, we turn to effect sizes for **w** < .1. (At **w** = 0, almost all universities would be unique.) The spread of the **w** values is much smaller than that of *z*: 97.6% of the comparisons result in **w** ≤ .1.[1] Whereas the maximum *z*-value in this data was between Harvard University and the University of Sao Paolo with *z* = 45.42, the largest **w-** value is only 0.41 based on the comparison between Nihon University and Rockefeller University. In other words, most of the differences between the universities seem negligible using this measure. In our opinion, these results by themselves would call into question the practice of producing rankings based on $PP_{top\ 10\%}$ since a ranking presumes meaningful differences.

[1] When the focus is on ***w***, node-sizes in the figures are determined by network characteristics using VOSviewer (Van Eck & Waltman, 2010).

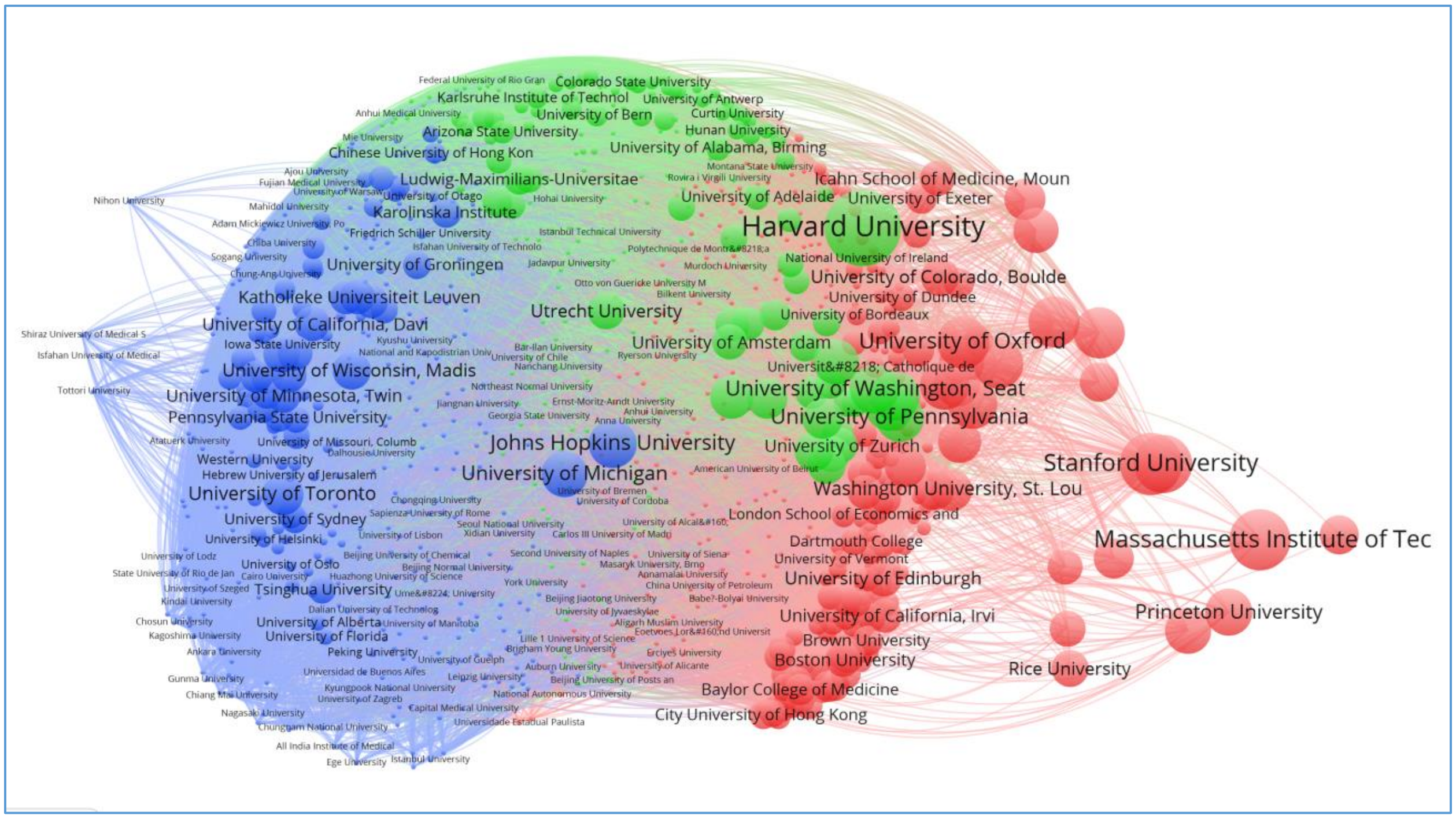

**Figure 4**: Three groups of universities are distinguished using ***w*** < 0.1; modularity $Q = 0.06$ (Blondel *et al*., 2008); layout VOSviewer.

Figure 4 shows three groups: on the right side of the figure the top universities are indicated in red.[2] In addition to American universities, some British universities are sorted into this class. Harvard University, however, is classified with a large group of universities in a second class indicated in green. As we noted in a study about university patenting (Leydesdorff, Etzkowitz, & Kushnir, 2016), the main divide perhaps emerges between a North-Atlantic (green) and an Asian-Pacific group (blue).

[2] Given the denser packings when using the much smaller **w-** values when compared with *z*-values, the modularity of the networks—indicated in the legends of the figures—is smaller by an order of magnitude.

When we raise the threshold to $\boldsymbol{w} < .3$, two groups remain (not shown here). The interpretation is again not obvious. It may be easier to provide the differences with an interpretation using national sets. We continue the discussion in the next sections using specific national systems of universities. One preliminary conclusion is that the effect sizes between universities are small even if the statistical significance of a difference is high. In other words, the two methods (of significance testing and effect sizes) do not indicate the same thing.

### *4.2. Universities in the United Kingdom*

The university system of the UK has been very much under discussion because of periodical evaluations such as the Research Excellence Framework (REF 2014; Wilsdon *et al*., 2015; Wouters *et al*., 2015). Although these evaluations are at the disciplinary level, they are organized institutionally at the university level. Universities are ranked in terms of a number between zero and five. These rankings have consequences for the funding.

LR 2017 covers 47 UK universities. Based on overlapping stability intervals among these universities, three groups are distinguished and one isolate, the London School of Hygiene and Tropical Medicine (Figure 5). Of the papers of this medical college, 21.1% belong to the top-10% set, against 18.4% for Oxford University which follows in the second position. The *z*-test generates two groups of universities at the 1% and three at the 5% significance level (Figure 6); with again the same isolate in both configurations. However, one should keep in mind that this is a representation of only 47 of the approximately 130 universities in the UK including former polytechnics and colleges. The latter group of (130 - 47 =) 83 universities can be considered as another group which is not included in the LR because of being insufficiently a research university.

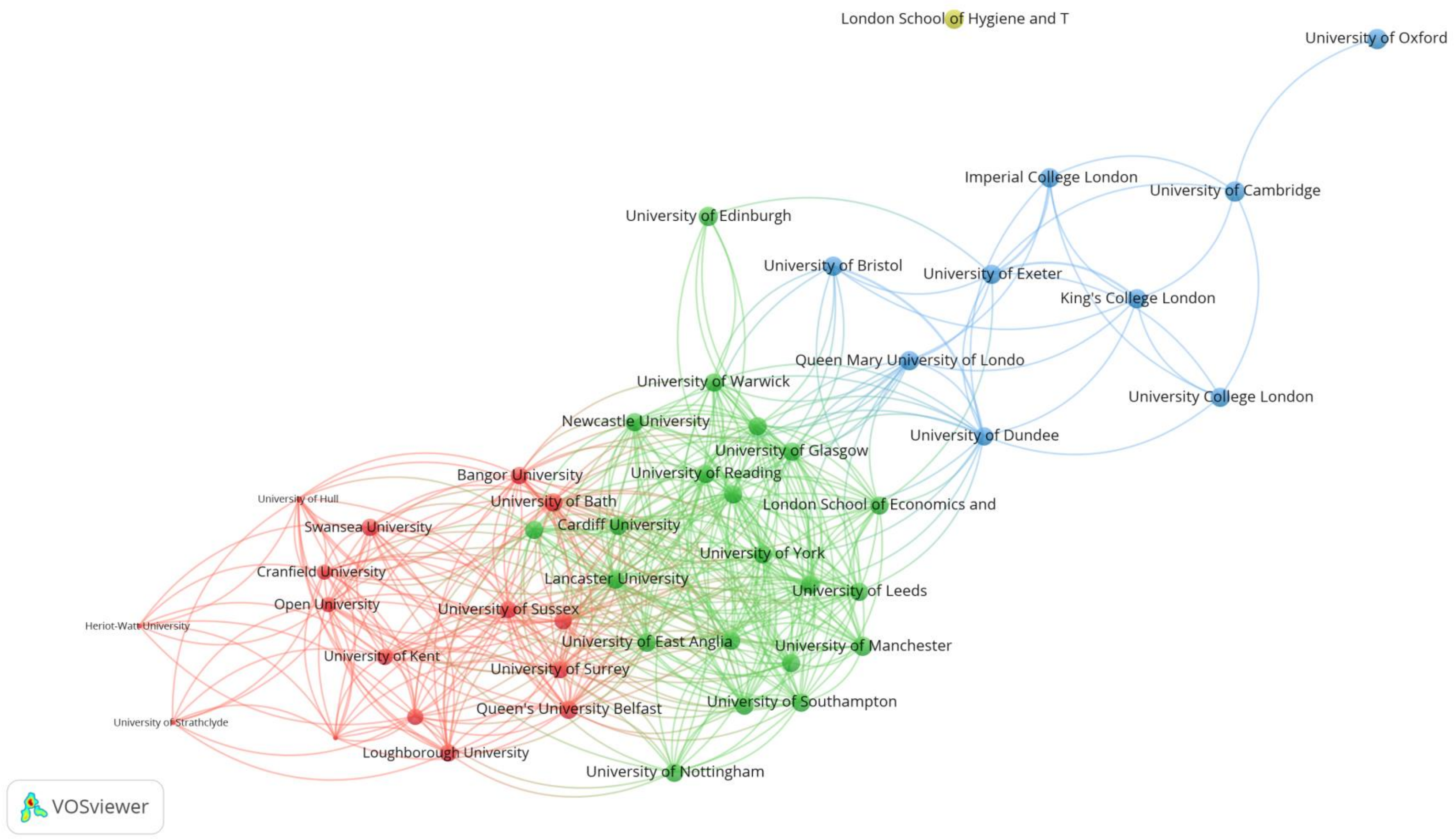

**Figure 5**: Overlapping stability intervals used as grouping criterion for 47 UK universities covered by the Leiden Ranking 2017; three communities distinguished with modularity $Q$ = 0.24 (Blondel *et al*., 2008); VOSviewer used for the visualization.

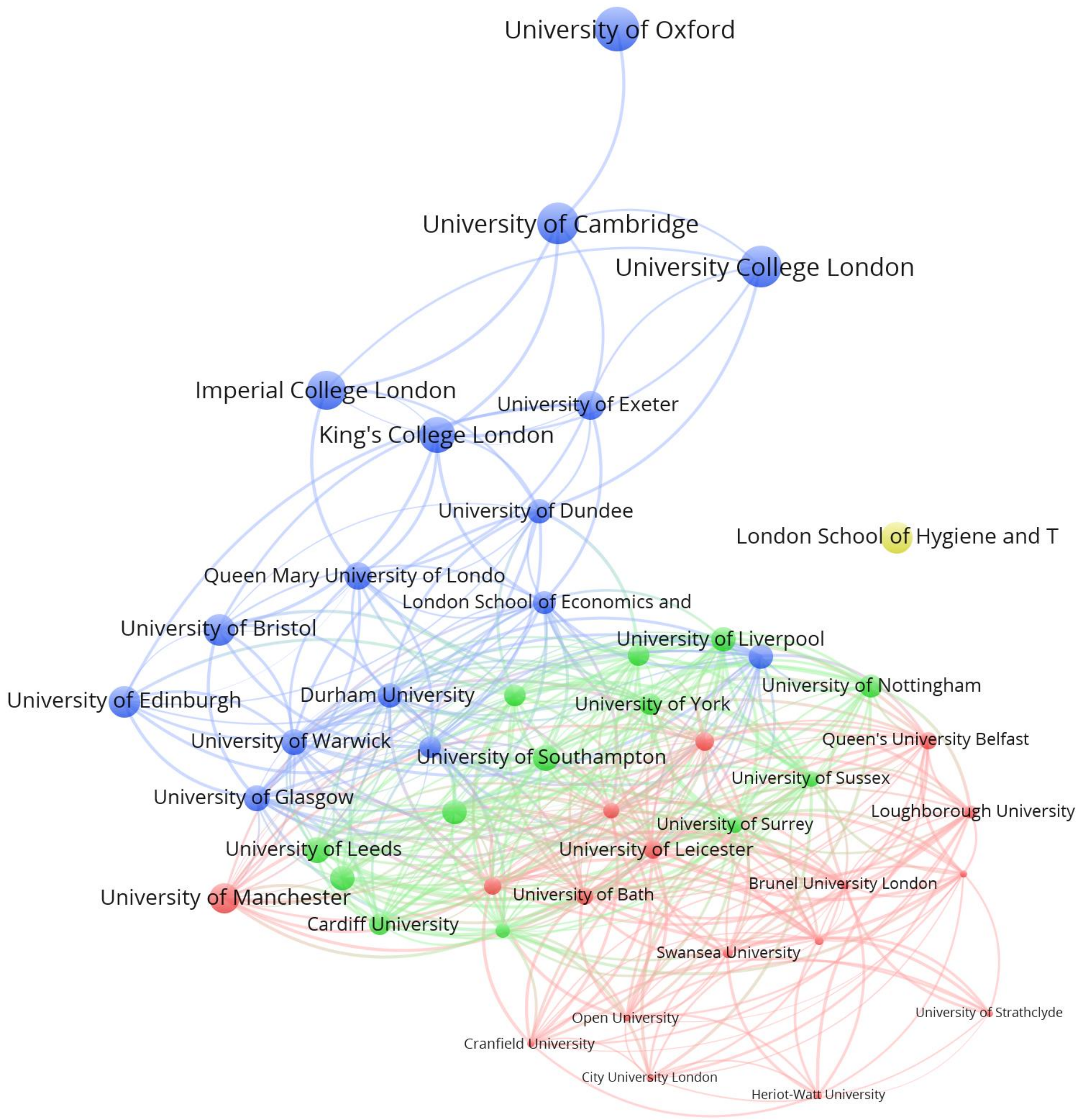

**Figure 6**: Classification of 47 universities in the UK; *p*<.05; *Q* = 0.20; layout VOSviewer.

Figure 6 shows the classification on the basis of significance testing at the 5% significance level. The figure is rather similar to Figure 5. Figure 7 shows the corresponding figure using effect sizes (*w* < .1). Two groups of universities are distinguished at both **w** < .1 and **w** < .3. (The corresponding figure for **w** < .3 is not essentially different.) However, the resulting classes are different from those based on statistical significance testing in important respects. Some universities belonging to the top-group in Figure 5 are now placed in the second group; for example, King's College London and Queen Mary University of London.

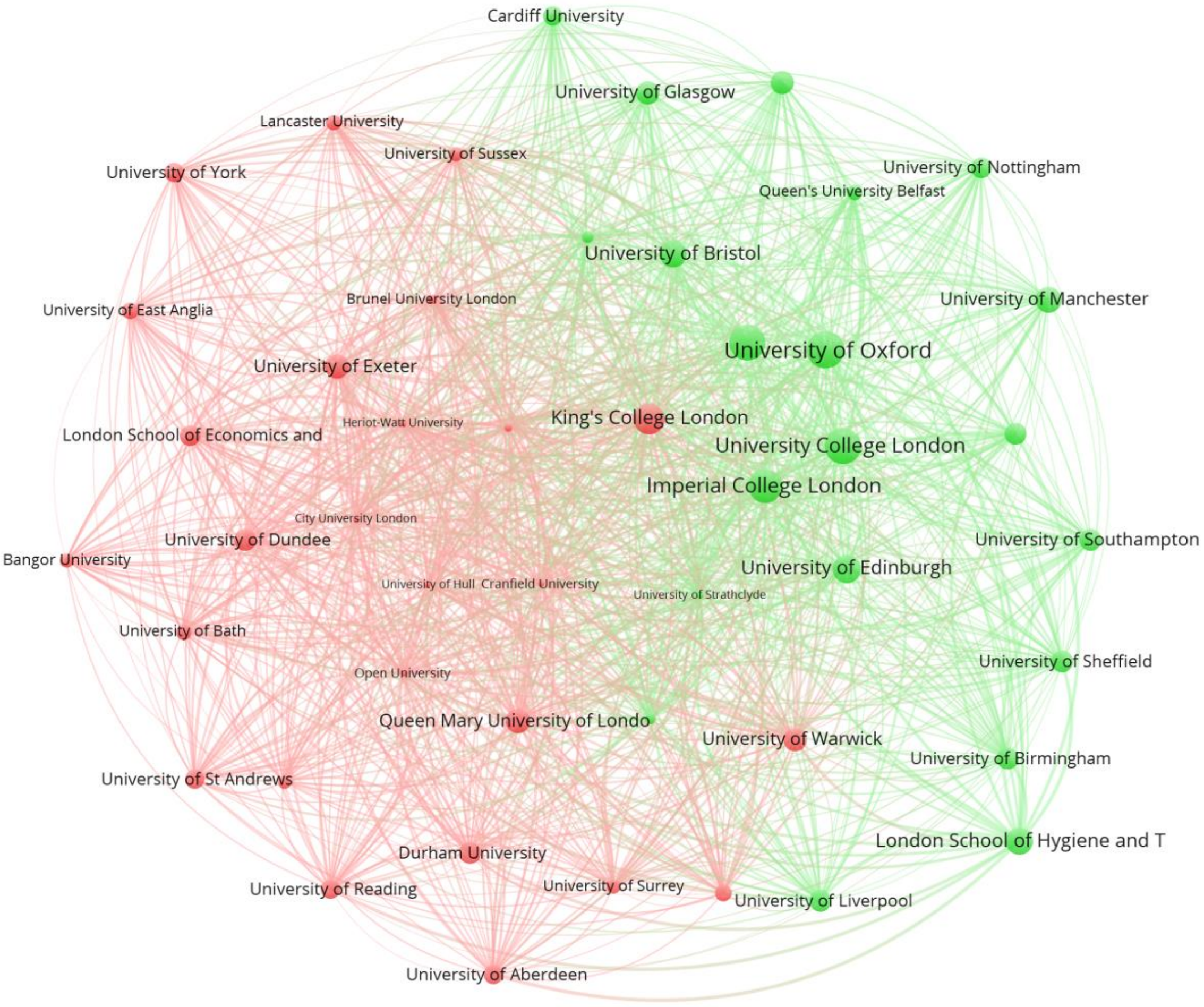

**Figure 7**: Classification of 47 UK universities in terms of effect sizes; **w** < .1; VOSviewer used for layout; *Q* = 0.024.

**Table 2:** Correspondence and differences among classifications of UK universities (*n* = 47).

| *Cramèr's V* | *Stability intervals* | *\|z\| < 2.576; p<.01* | *\|z\| < 1.96; p<.05* | *w<.1* |
|---|---|---|---|---|
| *p<.01* | 0.834*** | | | |
| *p<.05* | 0.869*** | 0.848*** | | |
| *w<.1* | 0.454 | 0.273 | 0.237 | |
| *w<.3* | 0.354 | 0.264 | 0.166 | 0.821*** |

***p < .001

A measure for the (lack of) correspondence between the classifications is provided by Cramèr's *V*, which is based on chi-square statistics, but which conveniently varies between zero and one. Table 2 shows these values among the five options discussed here: the classification based on overlapping stability intervals and the two classifications based on statistical significance testing, correlate highly ($V > 0.8$; $p < .001$); the two based on effect sizes also correlate ($V = .821$; $p < .001$); but there is a much lower correlation between classifications based on effect sizes and the other tests ($p > .05$). The relatively simple grouping on the basis of intersecting stability intervals is not outperformed by the other statistics.

**Table 3**: Group of universities on the UK's top-list indicated using different tests.

| *Overlapping stability intervals* | $\lvert z \rvert < 1.96$; *p <.05* | $\lvert z \rvert < 2.576$; *p <.01* | *w<.1* | *w<.3* |
|---|---|---|---|---|
| **Imperial College London** | **Imperial College London** | Bangor University | Cardiff University | Cardiff University |
| King's College London | King's College London | Durham University | **Imperial College London** | **Imperial College London** |
| Queen Mary University of London | Queen Mary University of London | **Imperial College London** | Loughborough University | King's College London |
| **University College London** | **University College London** | King's College London | Newcastle University | Loughborough University |
| **University of Bristol** | **University of Bristol** | London School of Economics and | Queen's University Belfast | Queen's University Belfast |
| **University of Cambridge** | **University of Cambridge** | Newcastle University | Swansea University | Swansea University |
| University of Dundee | University of Dundee | Queen Mary University of London | **University College London** | **University College London** |
| University of Exeter | University of Edinburgh | **University College London** | **University of Bristol** | University of Birmingham |
| **University of Oxford** | University of Exeter | University of Aberdeen | **University of Cambridge** | **University of Bristol** |
| | **University of Oxford** | **University of Bristol** | University of Edinburgh | **University of Cambridge** |
| | University of Reading | **University of Cambridge** | University of Glasgow | University of Edinburgh |
| | University of St Andrews | University of Dundee | University of Leeds | University of Glasgow |
| | | University of Edinburgh | University of Leicester | University of Leeds |
| | | University of Exeter | University of Manchester | University of Leicester |
| | | University of Leeds | University of Nottingham | University of Liverpool |
| | | University of Liverpool | **University of Oxford** | University of Manchester |
| | | **University of Oxford** | University of Southampton | University of Nottingham |
| | | University of Reading | University of Strathclyde | **University of Oxford** |
| | | University of St Andrews | University of Surrey | University of Sheffield |
| | | University of Surrey | | University of Southampton |
| | | University of Warwick | | University of Strathclyde |
| | | University of York | | |

Table 3 lists the UK universities that are on the top-list in each of the five classifications. Universities included on all five lists are boldfaced. Most of them are top universities as judged by the 2014 REF. However, on the basis of this most recent REF (bracketed figures are REF ranking):

1. King's College ($7^{th}$), LSE ($3^{rd}$) and Warwick ($9^{th}$) would be included in the boldfaced group;
2. Bangor ($42^{nd}$), Newcastle ($26^{th}$), Aberdeen ($46^{th}$), Dundee ($38^{th}$), Liverpool ($33^{rd}$), Reading ($39^{th}$), and Surrey ($45^{th}$) would not be in the top group.

Warwick is absent from three of the five listings in Table 3, whereas King's College London misses only on the list based on **w** <.1. The University of Edinburgh is not included when one uses overlaps of stability intervals as the criterion. The number of false positives is maximal (seven) for the classification based on $p$<.01 (but this is the longest list given this methodology). A choice between using effect sizes or significance testing is not obvious given these results, but the results are significantly different using either technique. It is also not obvious how the one analysis can inform the other. Again, the statistical significance tests suggest larger differences among universities than the analyses in terms of effect sizes.

*4.3 Germany*

The German science system has recently received very positive comments. "During a decade of global financial turbulence, her (that is, Angela Merkel's) government has increased annual science budgets in a stable, predictable, quintessentially German way. It has spurred competition among universities and improved collaboration with the country's unique publicly funded research institutions" (Abbott, 2017, at p. 18).

In 2006, the so-called Excellence Initiative was launched providing €1.9 billion of additional funding for three funding lines between 2006 and 2011: (1) graduate schools to promote early career researchers; (2) Clusters of Excellence to promote top-level research; and (3) institutional strategies to promote top-level university research (Bornmann, 2016). Universities awarded in the third funding scheme have been honoured with an elite status (Schröder *et al.*, 2014). As a result of the excellence initiative and further changes, according to Abbott (2017), "German universities have climbed up the world rankings. In 2005, only 9 German universities appeared in the Times Higher Education top 200. Now, there are 22. The LMU [Munich], which tops the German list in most years and has won in each round of the Excellence Initiative, rose from 61st place in 2011 to 30th in 2017" (p. 21).

Fifty German universities are included in the Leiden Ranking 2017. The *z*-test generates two groups of universities at the 1% and three at the 5% level. Figure 8 shows the latter three groups; universities which are in the excellence initiative of the German government are indicated by italicized labels in brown. (See Figure 1 above for the organization into two groups based on intersecting stability intervals.)

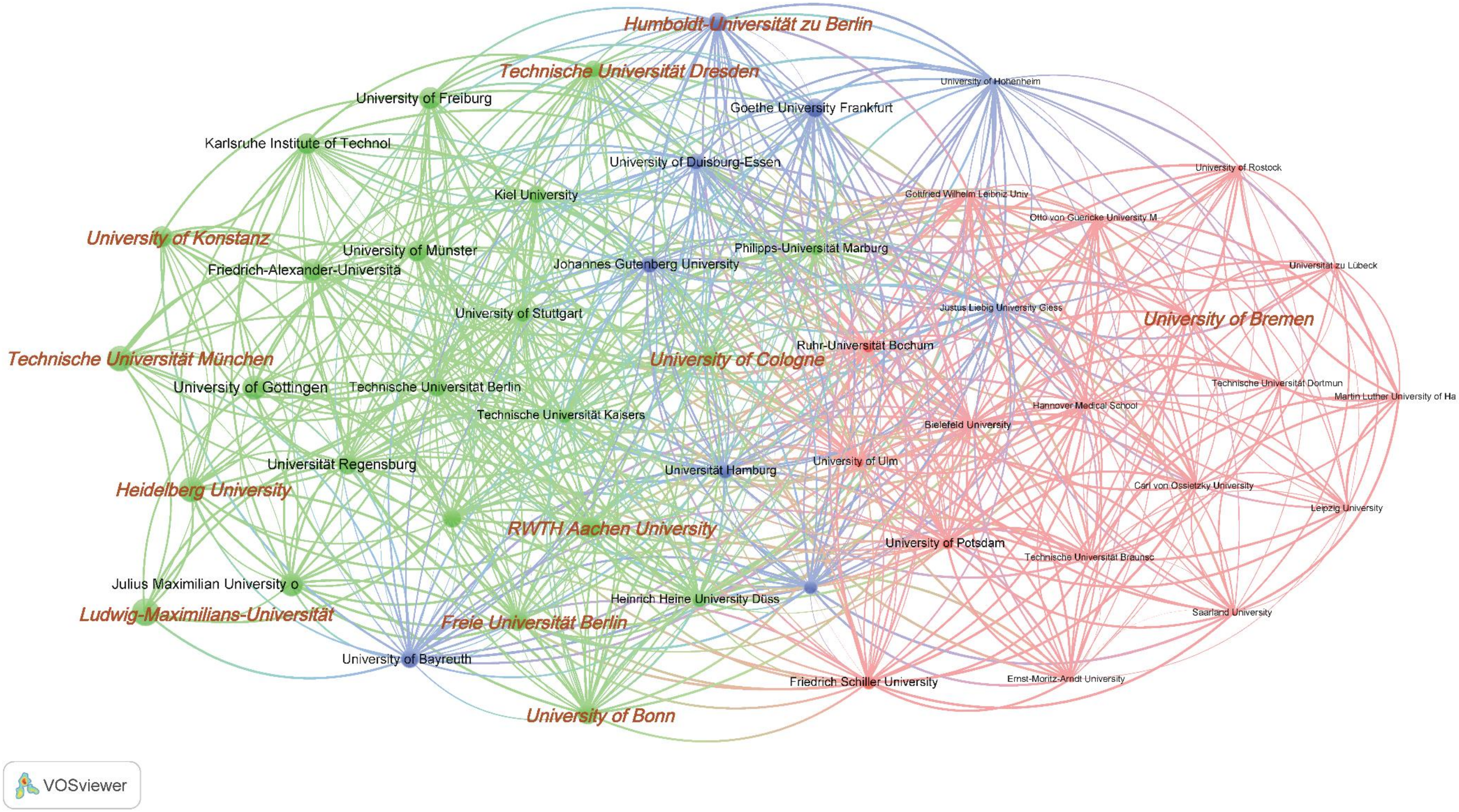

**Figure 8**: Classification of 50 research universities in Germany; *p* <.05; labels of excellent universities italicized and in brown.

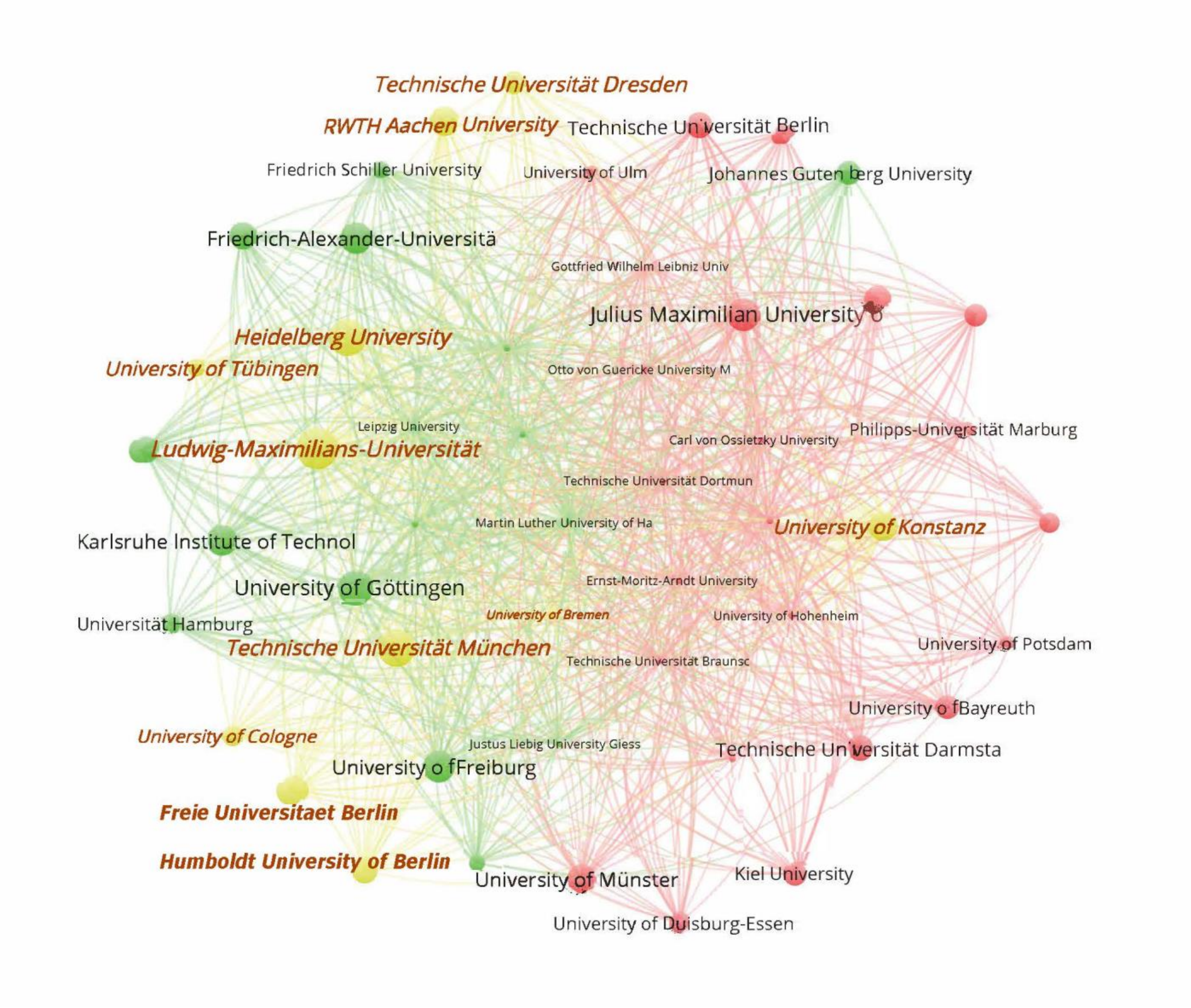

**Figure 9**: Classification of 50 research universities in Germany; **w** < 0.3

Figure 9 shows the classification using effect sizes. The difference between a map based on **w** < 0.3 or **w** < 0.1 is negligibly small because the **w**-values are anyhow smaller than 0.1 (with one exception).

Most universities that received grants in the Excellence Initiative are in the same group in both figures, but there are important exceptions: the University of Konstanz, for example, appears on the far left side in Figure 8, but on the right side in Figure 9. Conversely, the University of Bremen is part of the green group in Figure 9, but among the lower-ranked universities in Figure 8. The difference between these two universities—both belonging to the "excellence" group as defined by the Excellence Initiative—is statistically significant at the .001 level ($|z| = 4.252$). But the effect size **w** is still only 0.076 and therefore small. It is not within our purview to draw a policy conclusion or provide a recommendation other than noting this inconsistency between the selection of "excellence" by the bureaucracy and by LR 2017: one would not expect these two universities—Konstanz and Bremen—to be in the same class.

**Table 4:** Correspondence and difference among classifications of German universities ($n$ = 50).

| *Cramèr's V* | *Stability intervals* | *\|z\| < 2.576; p<.01* | *\|z\| < 1.96; p<.05* | *w<.1* |
|---|---|---|---|---|
| *p<.01* | 0.556*** | | | |
| *p<.05* | 0.825*** | 0.568*** | | |
| *w<.1* | 0.197 | 0.003 | 0.154 | |
| *w<.3* | 0.188 | 0.010 | 0.117 | 0.842*** |

***$p < .001$

The classifications can again be compared using Cramèr's *V* as we did above for the UK. The pattern is similar: effect sizes and statistical significance are two very different (orthogonal?) measures for testing differences. The choice of the statistical significance level has a larger effect in this case than in the case of the UK, but otherwise the results are similar. The UK results are less sensitive to parameter variations because the stratification among UK universities is more pronounced than in the German case.

In summary, German and UK universities are organized in not more than three classes: a top group, a middle one, and one at the bottom. However, even between members of different groups the effect sizes are not large. Any further fine-graining of the groups into subgroups or more specific rankings of individual universities is probably based on possible audience effects in the market as predicted by Gingras (2016, p. 75).

*4.4. Brazil*

Nineteen Brazilian universities are covered by the Leiden Ranking 2017. The scores of these universities are significantly below 10% on $PP_{top10\%}$ ($p < .001$). The largest effect size in a comparison ($w = 0.053$) is between the Federal University of Santa Catarina (UFSC) and the State University of Rio de Janeiro (UERJ), where the latter is at the bottom and the former at the top of the ranking with $PP_{top10\%} = 6.31$ and 3.72, respectively.

Using statistical significance testing, three groups of universities were distinguished; and using effect sizes or overlapping stability intervals, two. We asked Ricardo Sampaio, a Brazilian colleague who focuses on university rankings for domestic policy purposes, for comments. He noted that some universities are misplaced in the groupings (not shown as visuals here). For example, the Universidade Federal de Minas Gerais is considered as a top university in Brazil, but it is placed in the second class with $PP_{top10\%} = 5.11$. With $PP_{top10\%} = 5.41$, the University of Brasilia is also placed in this second group, but with $PP_{top10\%} = 5.71$ the University of Sao Paulo is ranked in the first grouping. In other words, the differences are small and seem more determined by the respective volume of publications than $PP_{top10\%}$. For example, the University of Sao Paulo produced 15,314 publications in the period under study (2012-2015), while this number is only 1,490 for the University of Brasilia.

### *4.5. United States*

Let us finally turn once more to the USA. As noted, U.S. universities dominated the patterns discussed in section 4.1 for the entire set. The figures for the 177 U.S. universities are not so different from what one would expect: a group of top universities (including Harvard, etc.), one or two medium groups, and a lower-ranked group. Figure 10a shows the four groups distinguished using overlapping stability intervals as criterion and Figure 10b the three groups on the basis of *z*-testing. In both cases and as before, MIT and Rockefeller University are additionally depicted as isolates.

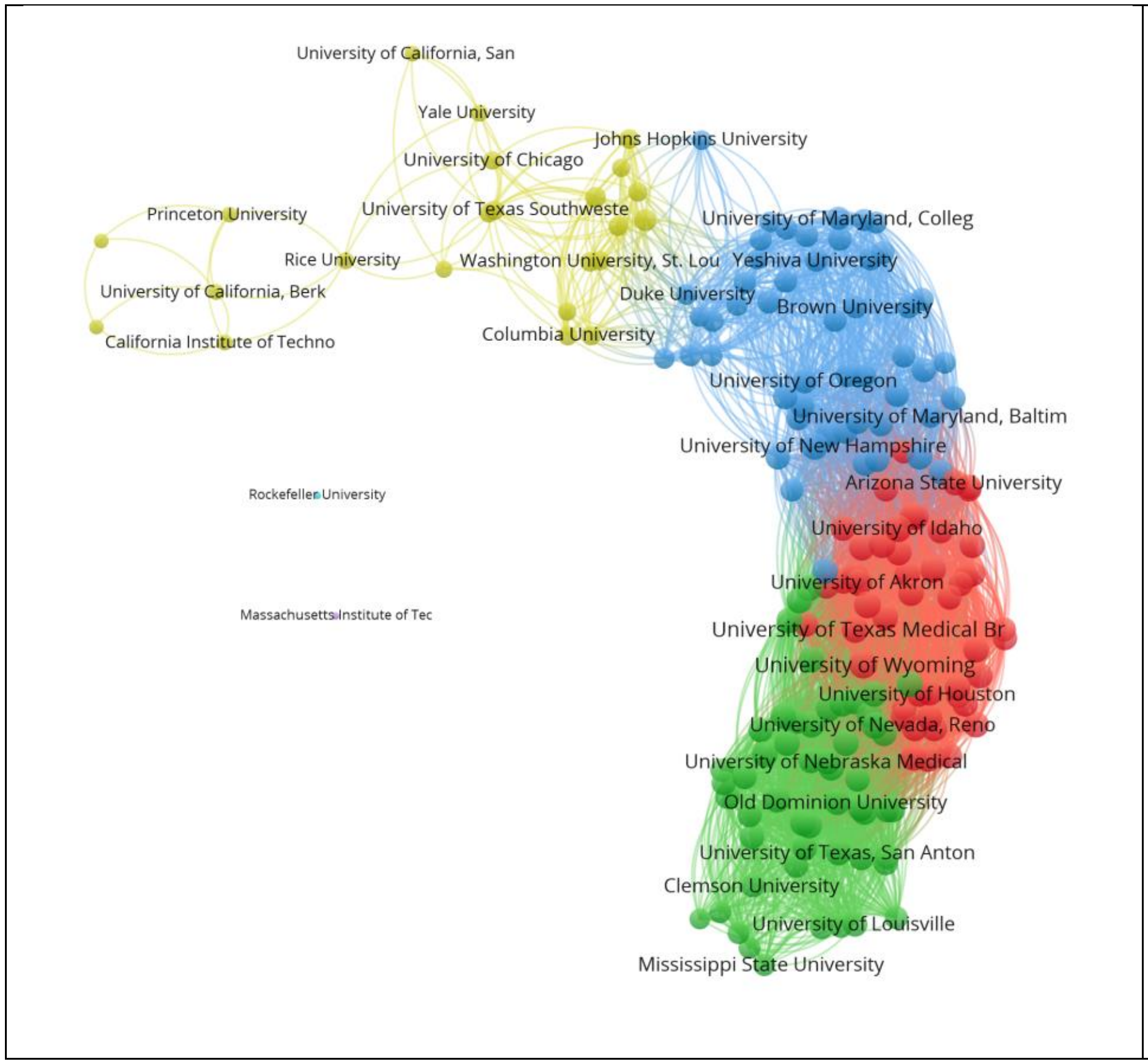

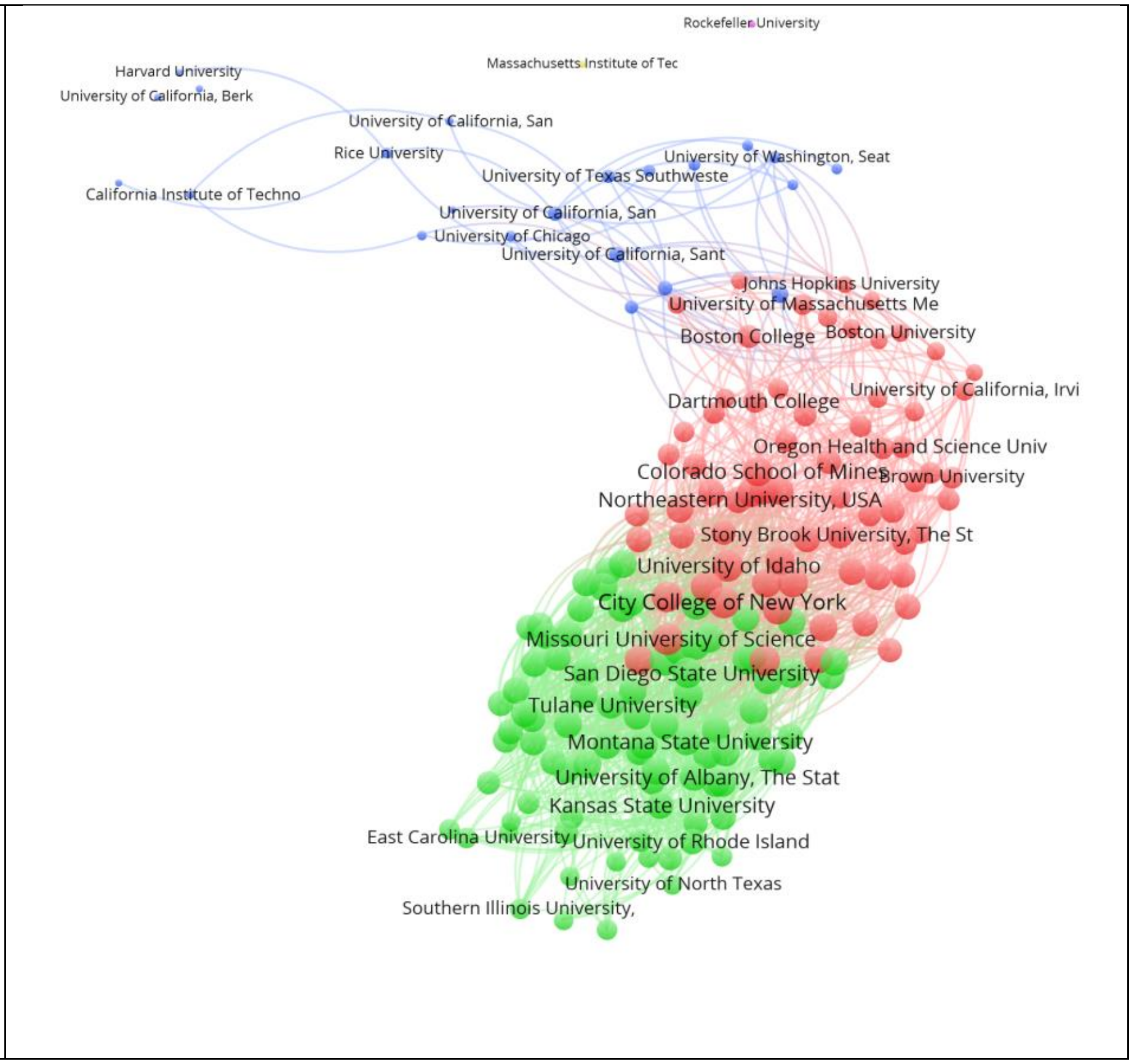

**Figures 10a and 10b:** 177 American universities clustered according to the criterion of overlapping stability intervals (six clusters including two isolates; $Q = 0.35$; Figure 10a on the left) and on the basis of signifcance testing ($p<.01$; $Q = 0.24$; five clusters including two isolates) in Figure 10b on the right.

If we change to effect sizes, two groups are distinguished using the threshold of ***w*** **<** 0.3 and three groups for ***w*** **<** 0.1. Figure 11 provides the solution for two groups. We have not been able to provide this major divide with a meaningful interpretation. Both major and less-known universities are present in both groups. Our previous suggestion to distinguish between an Atlantically and Pacifically oriented set does not hold at this level of the U.S. as a nation.

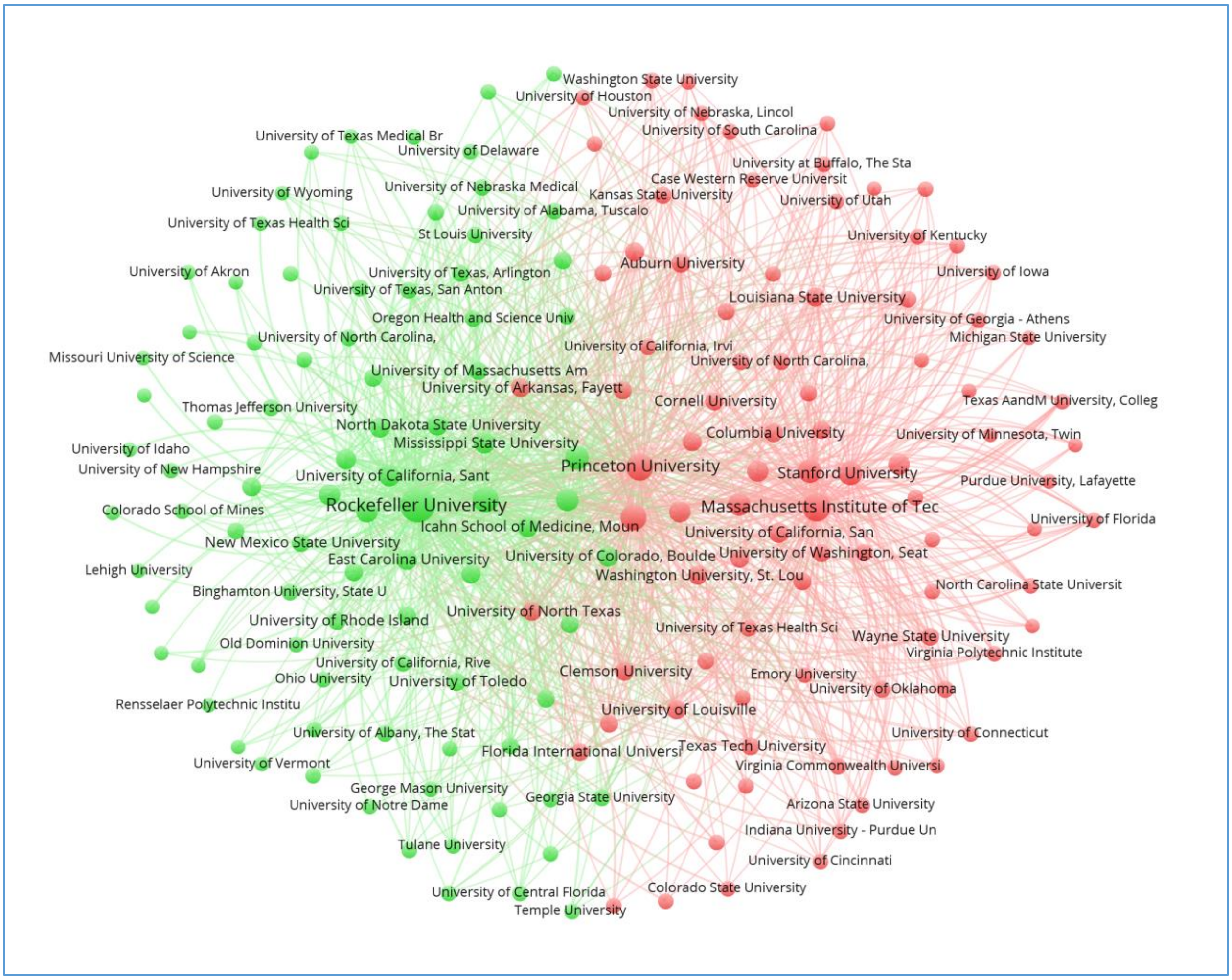

**Figure 11**: Distinction of two groups of American universities; *N* = 177; ***w*** < 0.3; VOSviewer used for the layout; Louvain algorithm for the decomposition.

**Table 5:** Correspondence and differences among classifications of 177 American universities.

| *Cramèr's V* | *Stability intervals* | *\|z\| < 2.576; p<.01* | *\|z\| < 1.96; p<.05* | *w<.1* |
|---|---|---|---|---|
| *p<.01* | 0.935*** | | | |
| *p<.05* | 0.831*** | 0.860*** | | |
| *w<.1* | 0.365*** | 0.487*** | 0.460*** | |
| *w<.3* | 0.343*** | 0.226 | 0.223 | 745*** |

***p < .001

Not surprisingly, Table 5 shows the same pattern as Tables 2 and 4 above. The larger sample, however, leads to more robust correlations.

*4.6. Comparison of effect sizes among national systems.*

Figure 12 shows the inequality among universities in these four national systems by plotting the effect sizes of the possible comparisons in decreasing order.

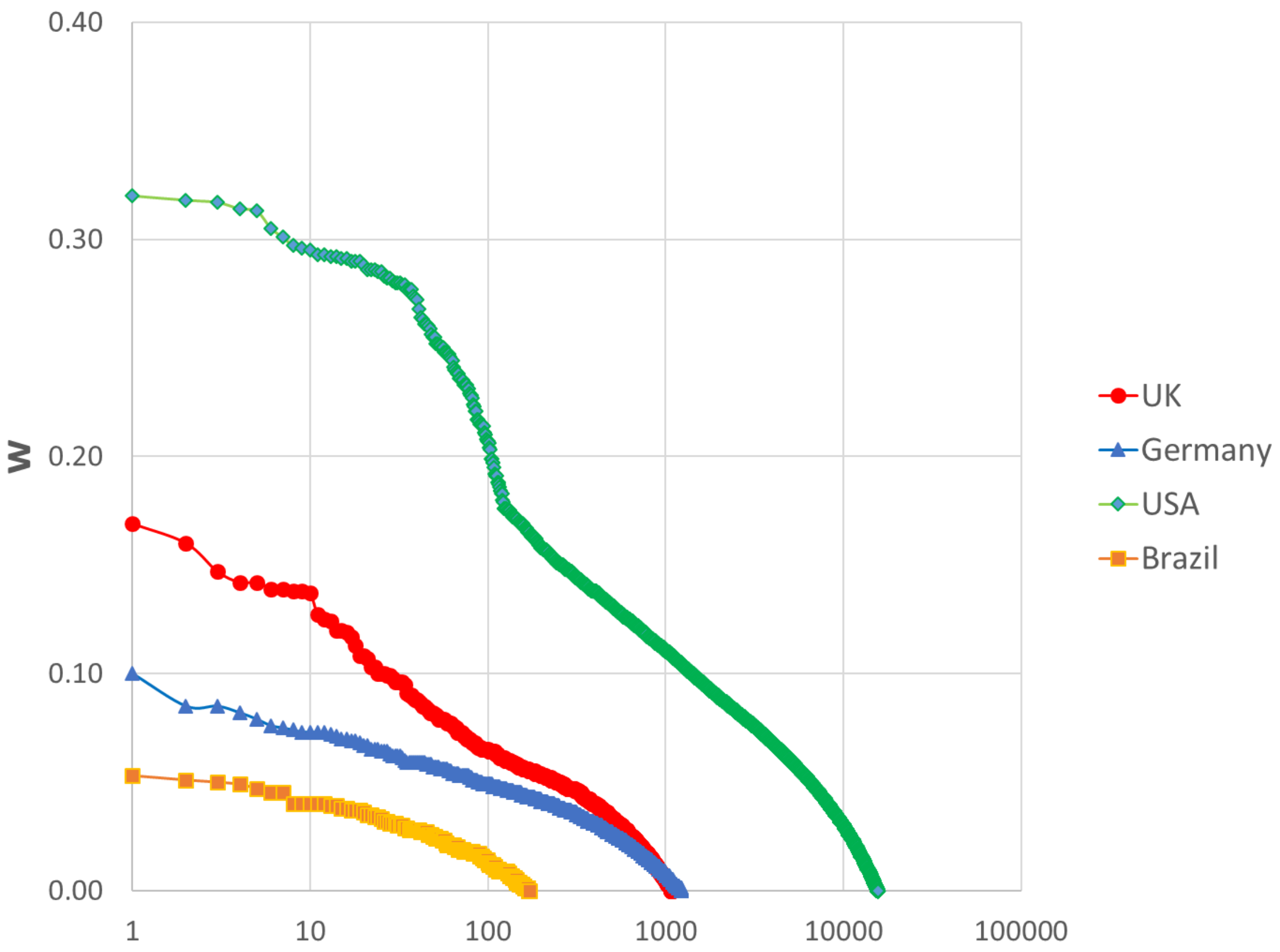

**Figure 12**: Distributions of (47 * 46 / 2 =) 1081 effect sizes **w** among 47 UK universities, (50 * 49 /2 = ) 1225 among 50 German universities, (177 * 176 /2 =) 15,576 among 177 American universities, and (19 * 18 /2) = 171 among 19 Brazilian universities.

American universities are most strongly stratified: **w** > .3 in seven comparisons and >.2 in 103 others. In the UK case, there are 26 comparisons (among 1081) with an effect size of **w** ≥ .1; in Germany, this effect size is virtually absent. The curves for Brazil and Germany are comparable, but at different levels.

## 5. Conclusions and discussion

We have analyzed the significance of differences in scores of universities on the LR 2017 in terms of effect sizes, stability intervals, and using the *z*-test. The main conclusion is that large groups of universities can be classified as belonging to the same group, and that differences among universities are often small if not negligible. Universities, in our opinion, tend to be isomorphic—that is, they operate under similar incentive structures and imitate one another. Both worldwide and at each country's level, a top-group can be identified and there is a further meaningful distinction between one or two groups in the middle range versus a group at the bottom. Note that there is also another group of universities which are not included in LR because they are not considered research universities or do not fullfil the requirements for inclusion.

Methodologically, our main conclusion is the unrelatedness of the differences using statistical significance tests or effect sizes. The results of the testing with ($z$-)statistics and stability intervals remain closer to the rankings and are intuitively more meaningful than the results of using effect sizes. The latter are not easily interpretable and sometimes counter-intuitive. Within each of the tests, parameter choices lead to relatively small changes in classifications. However, the measures themselves indicate different dimensions. We have not been able to provide the results in terms of effect sizes with a meaningful interpretation. Our results confirm the conclusion of Bornmann *et al.*'s (2013) analysis of LR 2011 that only 5% of the $PP_{top10\%}$ total variation can be traced back to differences between universities. Most of the variation can be explained by the location of universities in different countries.

## 6. Limitations

The classification suggests disambiguities whereas dividing lines may be much more fuzzy and polymorphic. The universities cannot be divided unambiguously, but each would have a range of possible ranks and associations depending on the reference sets and the parameters used by the analysts. We made parameter choices and used thresholds for effect sizes or *z*-values while knowing that there are no "bright-line" rules of yes/no decisions. However, the modularity algorithm imposes a clustering since a university on the border between two groups cannot be fractionally a member of both of them.

Furthermore, one can question the use of universities as units of analysis for rankings. It might be more appropriate to rank other units, such as research groups or departments. Universities are multi-disciplinary, whereas excellence is discipline- or even specialty-specific (Brewer *et al.*, 2001). A further limitation is the use of $PP_{top\ 10\%}$ as a specific indicator. We used the LR because of the quality of the data and the transparency of the methodology. Analogously to $PP_{top\ 10\%}$, we could have used $PP_{top\ 1\%}$, which is similarly available, or any other indicator in this ranking or another one. From a methodological perspective, $PP_{top\ 10\%}$ is a test-case. However, this indicator can also be considered as an "excellence indicator" (Bornmann, de Moya Anegón, & Leydesdorff, 2010; Leydesdorff, Wagner, & Bornmann, 2014).

Given these limitations, the main result is counter-intuitive and therefore interesting: cutting the sample into three or four groups may at first glance seem to ignore reality, but it is the only conclusion that we could draw. Although one should not reify these results, they provide an orientation. For practical purposes, our results suggest that networks based on overlapping stability intervals can provide a first impression of the relevant groupings among universities. The corresponding files (e.g., "Germany_o.net" underlying Fig. 1) can be read directly into a network analysis or visualization program.

## 7. Policy implications

The rankings generate a construct that seems to be fine-grained, but that can be analyzed as containing not more meaningful information than a division in three or four groups. The policy implication is that attempts to pursue rankings among universities focus on differences while the similarities and group structures of universities are backgrounded. Universities, however, are embedded in eco-systems, for example, at the national level. Competition among them has been induced by policies aiming to promote excellence. In the case of Germany, however, we found not always a direct relation between the Excellence Initiative of the German government and our grouping. Policies may be motivated also by other considerations than research excellence.

In the case of the UK, there were also important differences between the outcome of the REF 2014 and our classifications. We do not wish to claim priority for a statistical approach above a content-based one such as REF 2014 or the German Excellence Initiative. Discrepancies between the content-based and quantitative approaches may provide entrance points for further reflection and investigation. Our main aim has been to show that in each eco-system groups of universities are not significantly different. One may wish to differentiate policies for these different groups. However, we expect cultural patterns such as the prestige and status of a university to be sticky issues. One may have to fight an uphill battle to promote a peripheral university ahead of a central one.

**Acknowledgements**

We are grateful to Ricardo Sampaio, Ulrich Hoppe, and three anonymous referees for their comments.

**Appendix 1**

The measurement of effect sizes and the statistical significance of differences among universities using the Leiden Rankings.

1. Download the program from http://www.leydesdorff.net/software/leiden/leiden.exe .
2. Download the data at http://www.leidenranking.com/downloads in the Excel format.
3. Copy the fields "university," country," "field," "period," "fractional," "p," "p_top10," and "pp_top10" for the selection that one wishes to analyze to a separate worksheet; save this worksheet as "CSV (comma delimited)" to a file leiden.csv in the same folder as the program. Do not use another format (for Apple or DOS), since only this format preserves the diacritical characters.
4. Run the program; read the .net and .vec files into Pajek for the country under study. Within Pajek: > Options > Read – Write > UTF8;
5. Use Network > Transform > Remove > Lines > higher than > 1.96 for generating a file at the 5% level; *mutatis mutandis.*
6. Draw > (Network + Partition + Vector) > Export > 2D > VOSviewer. The vector file is needed for the node sizes.